\documentclass[prl, 11pt,letterpaper,superscriptaddress,floatfix,footinbib,notitlepage,reprint]{revtex4-1}

\usepackage{textcomp}
\usepackage{amsmath}
\usepackage{amsfonts}
\usepackage{amssymb}
\usepackage{graphicx}
\usepackage{siunitx}
\usepackage{caption}
\usepackage{setspace}
\setcitestyle{super}
\usepackage[colorlinks, citecolor={blue}, linkcolor={black}]{hyperref}

\begin{document}
\raggedbottom

\author{Yuan Cao}
\email{caoyuan@mit.edu}
\affiliation{Department of Physics, Massachusetts Institute of Technology, Cambridge, Massachusetts 02139, USA}
\author{Daniel Rodan-Legrain}
\affiliation{Department of Physics, Massachusetts Institute of Technology, Cambridge, Massachusetts 02139, USA}
\author{Jeong Min Park}
\affiliation{Department of Physics, Massachusetts Institute of Technology, Cambridge, Massachusetts 02139, USA}
\author{Fanqi Noah Yuan}
\affiliation{Department of Physics, Massachusetts Institute of Technology, Cambridge, Massachusetts 02139, USA}
\author{Kenji Watanabe}
\author{Takashi Taniguchi}
\affiliation{National Institute for Materials Science, Namiki 1-1, Tsukuba, Ibaraki 305-0044, Japan}
\author{Rafael M. Fernandes}
\affiliation{School of Physics and Astronomy, University of Minnesota, Minneapolis, MN 55455, USA}
\author{Liang Fu}
\affiliation{Department of Physics, Massachusetts Institute of Technology, Cambridge, Massachusetts 02139, USA}
\author{Pablo Jarillo-Herrero}
\email{pjarillo@mit.edu}
\affiliation{Department of Physics, Massachusetts Institute of Technology, Cambridge, Massachusetts 02139, USA}

\title{Nematicity and Competing Orders in Superconducting Magic-Angle Graphene}

\date{\today}

\begin{abstract}

\begin{singlespace}
Strongly interacting electrons in solid-state systems often display tendency towards multiple broken symmetries in the ground state. The complex interplay between different order parameters can give rise to a rich phase diagram. Here, we report on the identification of intertwined phases with broken rotational symmetry in magic-angle twisted bilayer graphene (TBG). Using transverse resistance measurements, we find a strongly anisotropic phase located in a `wedge' above the underdoped region of the superconducting dome. Upon crossing the superconducting dome, a reduction of the critical temperature is observed, similar to the behavior of certain cuprate superconductors. Furthermore, the superconducting state exhibits a anisotropic response to an directional-dependent in-plane magnetic field, revealing a nematic pairing state across the entire superconducting dome. These results indicate that nematic fluctuations might play an important role in the low-temperature phases of magic-angle TBG, and pave the way for using highly-tunable moir\'e superlattices to investigate intertwined phases in quantum materials.
\end{singlespace}

\end{abstract}

\maketitle

\section{Introduction}

Spontaneous symmetry breaking is a ubiquitous process that occurs at all length scales in nature\cite{anderson1972}, from the endowment of mass to elementary particles through the Higgs mechanism, the emergence of ferromagnetism and superconductivity in mesoscopic and macroscopic systems, all the way to the creation of stars and galaxies in the early universe. In a solid-state system, besides time-reversal and gauge symmetries, there are certain discrete symmetries imposed by the underlying crystal lattice. However, these symmetries can be spontaneously broken when many-body electron-electron interactions in the system are significant. Studying these broken-symmetry states is fundamental to elucidate the various phases in these many-body systems \cite{sachdev2003,keimer2015}. One example is an electronic nematic phase, where the discrete rotational symmetry of the lattice is spontaneously broken due to electron correlations, while lattice translational and time-reversal symmetries are preserved \cite{fradkin2010, fernandes2014}. The resulting anisotropy of the system is in turn manifested in the spin, charge, and lattice degrees of freedom, and can be measured via scattering, transport and scanning probe experiments. \cite{hinkov2008, chu2012, chuang2010, rosenthal2014, feldman2017, matsuda2017} 

When a correlated system has multiple broken-symmetry phases, their relationship often goes beyond mere competition, giving rise to a complex phase diagram of intertwined phases \cite{nie2014, tranquada2015,fernandes2019}. For example, in the underdoped region of the phase diagram of certain cuprate superconductors, a depletion in the critical temperature $T_c$ is found near $p\approx1/8$, where $p$ is the hole doping concentration \cite{taillefer2010}. This observation is typically attributed to the competition between superconductivity and a stripe phase that has spin and/or charge ordering \cite{taillefer2010,keimer2015,chang2012} Charge order and superconductivity may also intertwine to form a pair density-wave state \cite{tranquada2015,edkins2019}.
Another example of intertwined order is a nematic superconducting state, which simultaneously breaks lattice rotational and gauge symmetries. Nematic pairing states have been reported in certain iron pnictides and in doped Bi\textsubscript{2}Se\textsubscript{3}, as revealed by thermal, magnetic, and transport measurements \cite{li2017, shen2017, kuntsevich2018, matano2016, asaba2017, pan2016, smylie2018}, although their microscopic origin is still unclear.

The recent discovery of correlated insulator and superconducting behaviors \cite{cao2018,cao2018SC} in two-dimensional (2D) graphene superlattices brings the possibility of studying correlated superconducting materials with unprecedented tunability and richness. Twisted 2D materials exhibit long-range moir\'{e} patterns in real space that can be tuned by the twist angle (Fig. 1a). In twisted bilayer graphene (TBG) near the first magic-angle $\theta\approx\SI{1.1}{\degree}$, the interlayer hybridization results in nearly-flat bands at low energies, in which the electrons are localized in real space (Fig. 1a).\cite{morell2010, bistritzer2011, santos2012} Near half-filling of the nearly-flat bands, correlated insulator behavior and superconductivity have been demonstrated.\cite{cao2018, cao2018SC, yankowitz2019} These emergent states likely originate from strong electron-electron interactions in the nearly-flat bands. In this work we study the interplay between the superconducting phase and other many-body phases in magic-angle TBG. Compared to conventional materials, a major advantage of magic-angle TBG is that the band filling can be continuously tuned by electrostatic gating instead of chemical doping, so that different phases can be accessed in a single device.

\begin{figure*}[!ht]
\includegraphics[width=\textwidth]{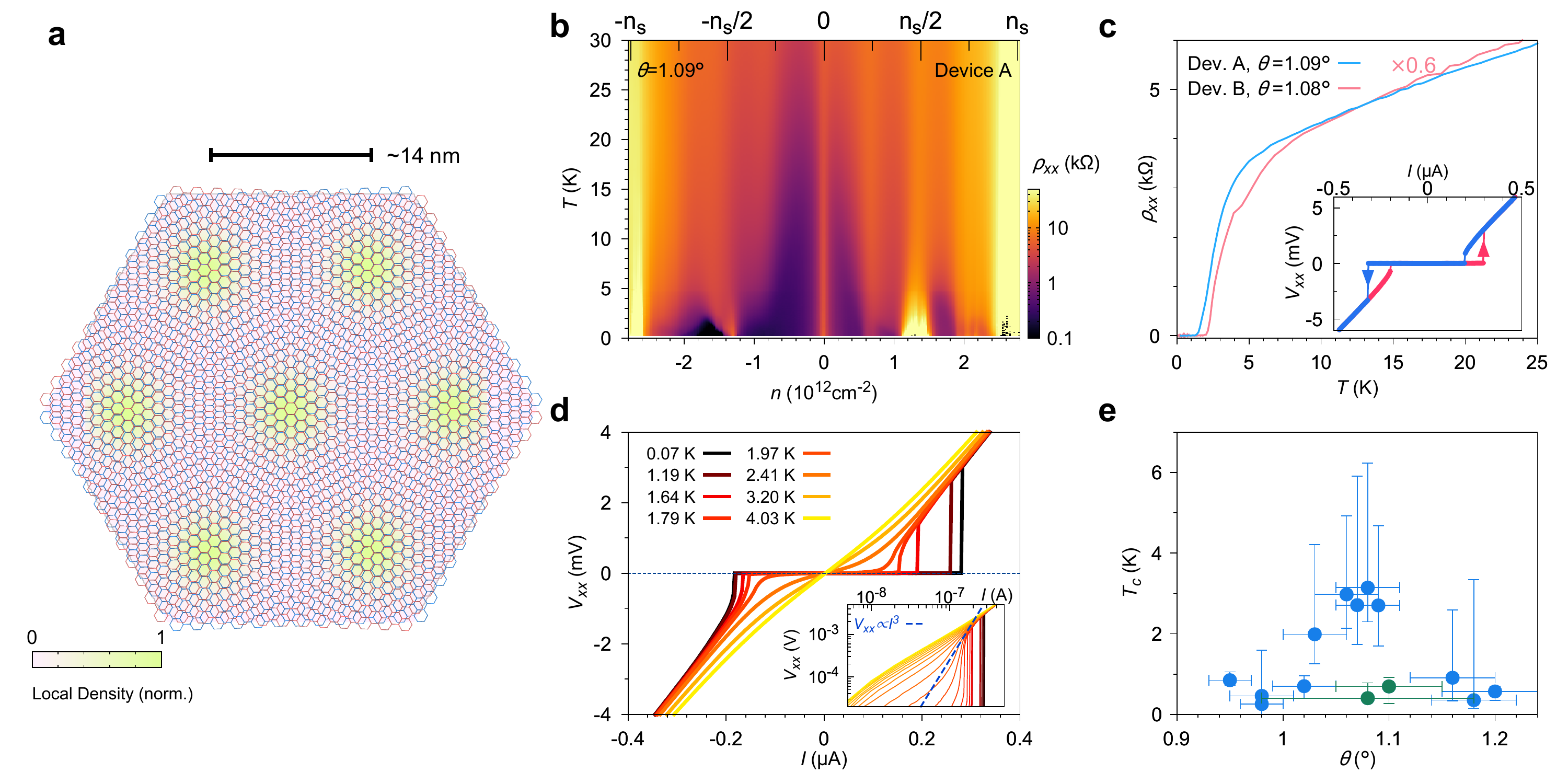}
\caption{\label{fig:fig1}(a) Illustration of the moir\'{e} pattern in twisted bilayer graphene (TBG). The color scale shows the normalized local density in the flat bands when the twist angle is close to magic angle. The twist angle of the displayed pattern is enlarged for clarity. (b) Resistivity of device A (twist angle $\theta=\SI{1.09}{\degree}$) versus gate induced carrier density and temperature, showing correlated features at all integer electron fillings of the superlattice. Superconductivity is found at hole-doping of the $-n_s/2$ insulator with critical temperature $\sim$\SI{2.5}{\kelvin}. (c) Resistivity versus temperature for devices A and B,  with twist angles $\theta=\SI{1.09}{\degree}$ and $\theta=\SI{1.08}{\degree}$, respectively, at their optimal doping concentrations. Inset shows the forward and backward sweeps of the $I$-$V_{xx}$ curves in device B which exhibit a significant hysteresis. (d) $I$-$V_{xx}$ curves at different temperatures measured in device B. Inset shows the log-log plot of the $I>0$ part of the data. The Berezinskii-Kosterlitz-Thouless transition temperature $T_{\mathrm{BKT}} \approx \SI{2.2}{\kelvin}$ is identified where the slope of the curve crosses $\mathrm{d}(\log V_{xx})/\mathrm{d}(\log I) = 3$ (equivalent to $V_{xx}\propto I^3$). (e) Statistics of optimal doping $T_c$ in 14 of the magic-angle TBG devices we have measured. We find that the trend of $T_c$ peaks around \SI{1.1}{\degree}, the theoretically predicted first magic-angle in TBG. The green data points are from devices exhibiting substantial disorder, hence the large error bars in the twist angle determination. This disorder may be responsible for the relatively low $T_c$.}
\end{figure*}

\section{Results}

In this article, we investigate the phase diagram of magic-angle TBG in detail, focusing particularly on anisotropic properties in the superconducting and normal phases. We uncover an anisotropic in-plane electrical transport in magic-angle TBG at low temperatures using longitudinal and transverse resistivity measurements. In addition, we reveal an anisotropic in-plane critical field and an anistropic response of the superconducting critical current to an in-plane magnetic field. Our results show that magic-angle TBG can spontaneously break lattice rotational symmetry in both the normal and superconducting phases, although the anisotropic properties of these two states are manifested in different observables, suggesting that the origins of these two anisotropic states might be different. 

\subsection{Characterization of Magic-angle Graphene}

Using the previously developed `tear and stack' dry-transfer technique \cite{cao2016, kim2016}, we fabricate high quality encapsulated TBG devices with twist angles around the first magic angle $\theta\approx\SI{1.1}{\degree}$. The main devices we report about are devices A and B, with twist angles of $\theta=\SI{1.09}{\degree}$ and $\theta=\SI{1.08}{\degree}$, respectively. The low-energy bands in TBG are four-fold degenerate (due to spin and valley degrees of freedom) and can sustain an electron density of $n_s=4/A$, where $A$ is the area of a moir\'{e} unit cell. This density corresponds to filling four electrons or holes per moir\'{e} unit cell. Near the first magic angle, correlated states can form at integer electron fillings of the moir\'e superlattice, \emph{i.e.} when $n=\pm\frac{n_s}{4}, \pm\frac{n_s}{2}, \pm\frac{3n_s}{4}$. This is believed to be a consequence of the fact that the electronic interactions become comparable to the bandwidth of the nearly-flat bands. In the resistivity measurements of device A shown in Fig. 1b, we indeed find an enhancement of the resistivity $\rho_{xx}$ at all these integer fillings. A superconducting dome is recognizable upon hole-doping of the $-n_s/2$ insulating state, at temperatures below \SI{2.5}{\kelvin}. Fig. 1c shows the $\rho_{xx}(T)$ curves of device A and device B at their optimal doping levels (highest $T_c$). Both devices exhibit a relatively high $T_c$ in the range of \SIrange{2.5}{3}{\kelvin}  (at \SI{50}{\percent} normal resistance) \cite{supplementary}. Figure \ref{fig:fig1}d shows the evolution of the $I$-$V$ curves with temperature. From the log-log plot shown in the inset, we can extract the Berezinskii-Kosterlitz-Thouless (BKT) transition temperature to be $T_\mathrm{BKT}\approx \SI{2.2}{\kelvin}$. Devices A and B have in fact some of the highest transition temperatures among all reported magic-angle TBG devices so far, as evident from the $T_c$ statistics shown in Fig. 1e, as well as devices reported in the literature \cite{cao2018SC, yankowitz2019, lu2019}. 

\begin{figure*}[!ht]
\includegraphics[width=\textwidth]{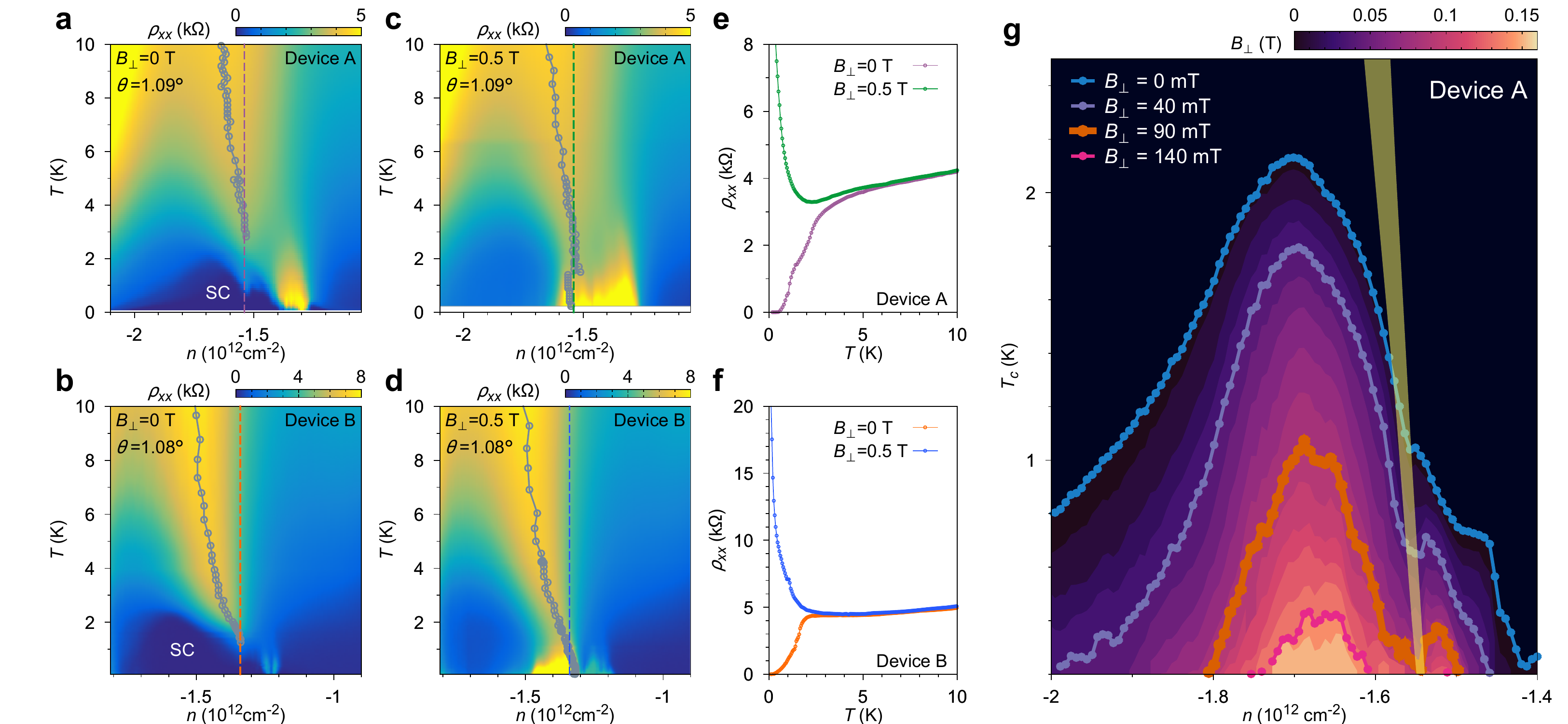}
\caption{\label{fig:fig2} Competing phases near the superconducting domes of magic-angle TBG. (a-b) Resistivity versus gate-induced carrier density and temperature for device A and device B, respectively.  (c-d) Same measurement but in a perpendicular magnetic field of \SI{0.5}{\tesla}. (e-f) Line cuts of resistivity versus temperature for devices A and B at \SI{0}{\tesla} and \SI{0.5}{\tesla} at the densities indicated by the dashed lines in (a-d), showing a superconductor-to-insulator transition induced by the magnetic field. In both devices, we find a wedge-like feature above the superconducting dome in addition to the $-n_s/2$ correlated state. This feature shifts noticeably towards negative relative densities at elevated temperatures. At zero magnetic field, the wedge-like feature disappears into the superconductiviting dome and creates a `kink' on the $T_c(n)$ curve, while in $\SI{0.5}{\tesla}$ it persists down to the lowest temperatures and turns into an insulator. The gray circles in (a-d) indicate the resistivity maxima associated with the wedge-like feature at different temperatures. (g) Evolution of $T_c$ of device A in a perpendicular magnetic field up to \SI{0.18}{\tesla}. Each contour line is $T_c$ (\SI{10}{\percent} normal resistance) versus carrier density at the magnetic field indicated by the color shading. The yellow band represents the approximate position of the wedge-like feature. At $B_\perp=\SI{0.09}{\tesla}$ (orange curve), the $T_c$-kink touches zero at the same density where the wedge-like feature extrapolates to zero temperature. Above this field, the superconducting dome splits into two domes roughly centered at $\SI{-1.52e12}{\per\centi\meter\squared}$ and $\SI{-1.67e12}{\per\centi\meter\squared}$, respectively.
}
\end{figure*}

\subsection{Anisotropic Behavior in the Normal Phase}

Figs. 2a-b show the resistivity versus gate-induced density, $n$, and temperature, $T$, maps of devices A and B, respectively, in the vicinity of $-n_s/2$. We find that in both devices the $-n_s/2$ region of the phase diagram has a rather complicated structure. As can be seen in Figs. 2a-b, there are two resistive features in the normal state: one `wedge'-like feature above the superconducting dome (near \SI{-1.5e12}{\per\centi\meter\squared} for device A and \SI{-1.4e12}{\per\centi\meter\squared} for device B) that bends at elevated temperatures, and one resistive feature on the right hand side of the dome (near \SI{-1.3e12}{\per\centi\meter\squared} for both devices). While the latter feature corresponds to the $-n_s/2$ state similar to the correlated states previously reported in magic-angle TBG \cite{cao2018, cao2018SC, yankowitz2019, lu2019}, the wedge-like feature creates a noticeable `kink' (\emph{i.e.} decrease in $T_c$) where it intersects with the superconducting dome. This suppression of $T_c$ resembles that observed in underdoped cuprates, where it is attributed to a spin/charge ordered phase that competes with superconductivity \cite{taillefer2010,keimer2015,chang2012}. To further probe the resistive wedge-like feature, we apply a small perpendicular magnetic field to fully suppress superconductivity, as shown in Figs. 2c-d. Line cuts of the resistivity versus temperature at the densities corresponding to the `kinks' of $T_c$ are compared in Figs. 2e-f for the two devices. It can be clearly seen that when superconductivity is suppressed, the resistive wedge-like feature turns insulating upon approaching zero temperature. A small magnetic field thus results in a superconductor-to-insulator transition at this density. In Fig. 2g, we show the gradual suppression of $T_c$ by the perpendicular magnetic field from zero to \SI{180}{\milli\tesla} in device A. We find that above about \SI{90}{\milli\tesla}, the superconducting dome splits at $n\approx \SI{-1.54e12}{\per\centi\meter\squared}$ into two domes. This density approximately coincides with the density where the wedge-like feature extrapolates to zero temperature. The separated domes are centered at around $\SI{-1.52e12}{\per\centi\meter\squared}$ and $\SI{-1.67e12}{\per\centi\meter\squared}$ respectively.  The position of the splitting point corresponds to \SI{15+-5}{\percent} hole doping with respect to the correlated insulator state. These findings are reminiscent of the recently reported high-field experiments in underdoped cuprates, where the superconducting dome splits at $1/8$ hole-doping in a magnetic field $>\SI{30}{\tesla}$ \cite{ramshaw2015}, suggesting that a quantum critical point might also exist in the phase diagram of magic-angle TBG.

\begin{figure*}[!ht]
\includegraphics[width=\textwidth]{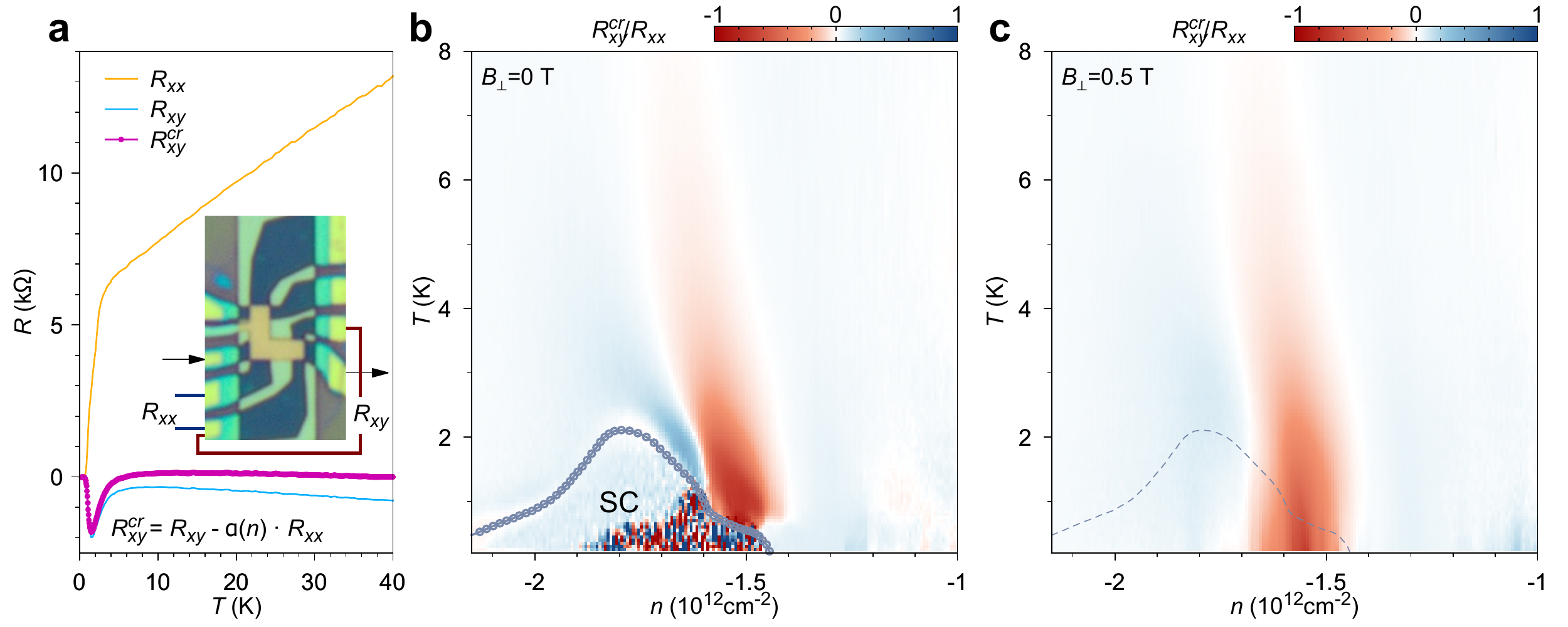}
\caption{\label{fig:fig3} Normal-state anisotropy in magic-angle TBG device A. (a) Illustration of how we extract the resistivity anisotropy by measuring the transverse resistance. Inset shows the actual device configuration that we used to obtain the data, where black arrows indicate current source and drain. $R_{xx}$, $R_{xy}$ label the leads on which longitudinal and transverse resistances are measured, respectively. The purple trace, $R_{xy}^\mathrm{cr}$, is the transverse resistance corrected for leads misalignment.\cite{supplementary} $n=\SI{-1.53e12}{\per\centi\meter\squared}$ in this measurement. (b-c) Anisotropy ratio versus carrier density and temperature at zero magnetic field and $B_\perp=\SI{0.5}{\tesla}$, respectively.\cite{supplementary} The circles in (b) and the dashed line in (c) outline the approximate shape of the superconducting dome (at zero field). We find the strongest anisotropy near the kink in $T_c$ at $n=\SI{-1.54e12}{\per\centi\meter\squared}$, coinciding with the resistive wedge-like feature we identified in Figs. 2a and 2g.}
\end{figure*}

To gain more insight into the possible origin of the resistive wedge-like feature, we measure the transverse voltage across the sample at zero magnetic field, which gives us the transverse resistance $R_{xy} = V_y/I_x$ \cite{walmsley2017, wu2017}. In an anisotropic conductor in two dimensions, the 2-by-2 resistivity tensor has two diagonal components $\hat{\rho}=\mathrm{diag}\{\rho_1, \rho_2\}$. If the major axis of the anisotropy (usually one of the crystal axis) is not aligned with the reference frame of the tensor, the off-diagonal terms of the resistivity tensor are proportional to $(\rho_1-\rho_2)\sin(2\theta)$, where $\theta$ is the angle between the anisotropy axis and the reference $x$-axis (see Supplementary\cite{supplementary} for derivation).  As a result, when an electrical current $I_x$ flows in the $x$ direction, a transverse voltage $V_y$ appears across the edges perpendicular to the $y$ axis, giving a nonvanishing $R_{xy} = V_y/I_x \propto (\rho_1-\rho_2)\sin(2\theta)$ as long as $\sin(2\theta)\neq 0$ and $\rho_1 \neq \rho_2$. The first condition is assumed to be true in our experiment, since the lattice orientation is random with respect to the sample edge. Consequently, a non-vanishing transverse resistance in our experiment implies anisotropic resistivity, $\rho_1 \neq \rho_2$, and therefore the breaking of the six-fold rotational symmetry of TBG. Note that this transverse voltage is fundamentally different from the Hall effect since time-reversal symmetry is not broken. In order to quantitatively analyze the transverse voltage, we need to remove any residual longitudinal component that might appear in the transverse voltage due to imperfect alignment of the four-probe voltage contacts and/or sample inhomogeneity.\cite{walmsley2017,supplementary} Fig. 3a shows the raw $R_{xx}$ and $R_{xy}$ measured for device A near the wedge-like feature, as shown in Fig. 2a. At high temperatures (\SI{40}{\kelvin}), where the anisotropies associated with electron correlation effects are presumably overwhelmed by thermal fluctuations, both $R_{xx}$ and $R_{xy}$ are linear in $T$ and proportional to each other: $R_{xy} \approx -0.05 R_{xx}$.\cite{walmsley2017} To correct for this background signal that is likely a result of the imperfect voltage probe alignment, we subtract this $R_{xx}$ component from $R_{xy}$ so that at the highest temperature of \SI{40}{\kelvin} the net signal is zero. This corrected transverse voltage $R_{xy}^{\mathrm{cr}} = R_{xy} - \alpha(n)R_{xx}$, where $\alpha(n)$ is a density-dependent numerical factor typically within $\pm0.1$, constitutes a truthful measure of the resistivity anisotropy (purple curve in Fig. 3a). We note that, while no signal is present at higher temperatures, below $\SI{6}{\kelvin}$ there is a significant negative peak in $R_{xy}^{\mathrm{cr}}$, which indicates the onset of anisotropy at this temperature.

The gate and temperature dependence of the anisotropy, shown in Figs. 3b-c for zero magnetic field and $B_\perp=\SI{0.5}{\tesla}$ (see Supplementary\cite{supplementary}), clearly reveals a prominent anisotropy `wedge' as well. The transverse voltage measured at $B_\perp=\SI{0.5}{\tesla}$ is symmetrized with data measured at $B_\perp=-\SI{0.5}{\tesla}$ to remove the contribution from the Hall voltage. Here we plot the normalized quantity $R_{xy}^\mathrm{cr}/R_{xx}$, which is approximately proportional to the anisotropy ratio $\frac{\rho_1-\rho_2}{\rho_1+\rho_2}$ (see Supplementary \cite{supplementary}). We also mark out the superconducting dome in Fig. 3b-c. Immediately above the superconducting dome on the `underdoped' side (lower $|n|$), we find a strong transverse voltage signal with a sign change at around \SI{-1.59e12}{\per\centi\meter\squared} (see the supplementary for other ranges of density).\cite{supplementary} The position of the anisotropy wedge matches well with the resistive wedge-like feature that we observed in Fig. 2a. The sign change indicates that the anisotropy changes from $\rho_1>\rho_2$ to $\rho_1<\rho_2$ (or vice versa). In $B_\perp=\SI{0.5}{\tesla}$ (Fig. 3c), the anisotropy wedge with negative values of $R_{xy}^\mathrm{cr}$ persists to zero temperature, consistent with the behavior of the resistive wedge-like feature in Fig. 2b as well. On the other hand, we notice that the anisotropy with positive $R_{xy}^\mathrm{cr}$ near $\SI{-1.65e12}{\per\centi\meter\squared}$ disappears as superconductivity is suppressed by the magnetic field, which might be explained by the vestigial order from the nematic superconductivity that will be discussed in the next section.

\begin{figure*}[!ht]
\includegraphics[width=\textwidth]{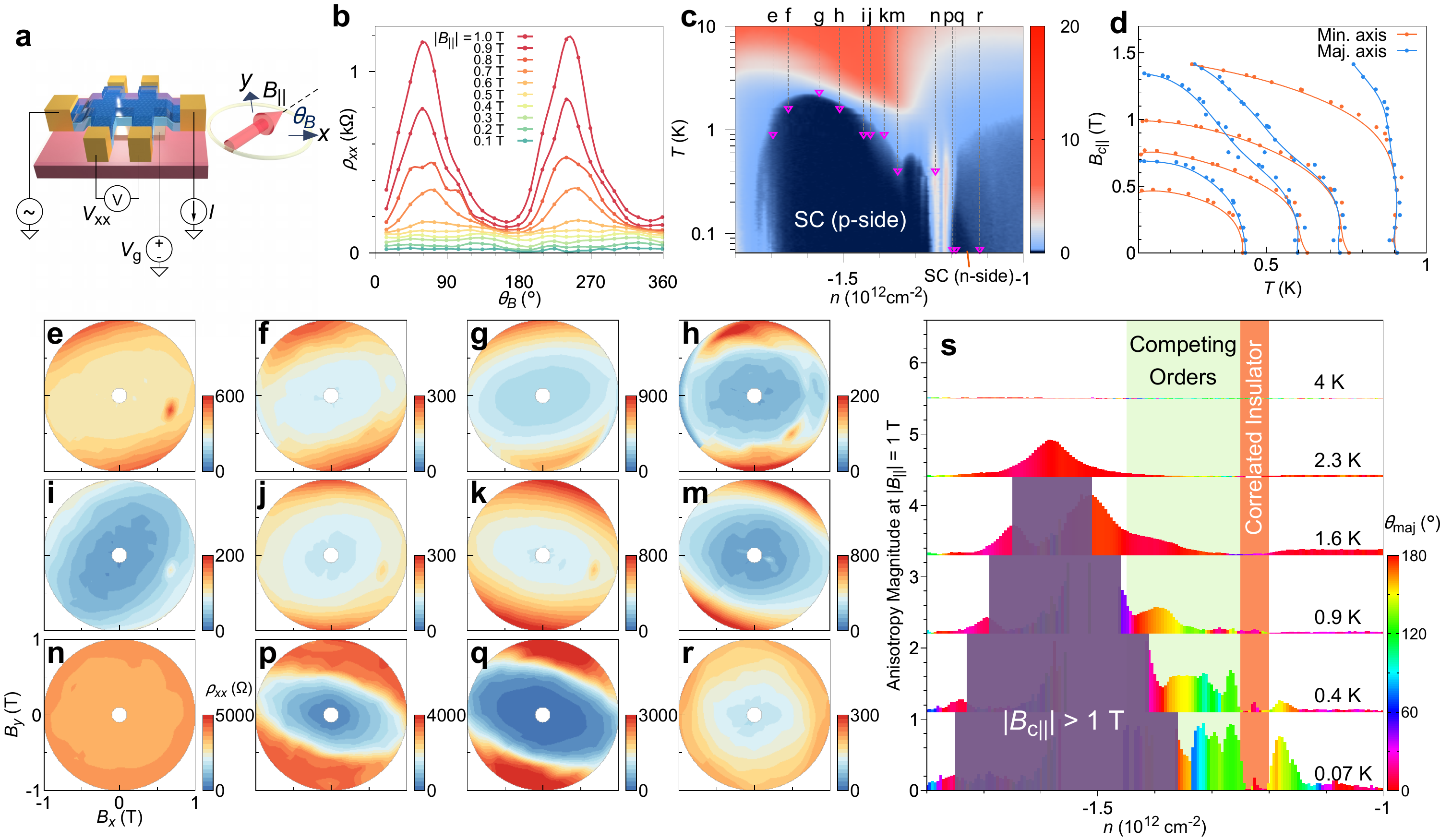}
\caption{\label{fig:fig4} Evidence for nematic superconductivity in magic-angle TBG. (a) Definition of the in-plane field angle $\theta_B$ with respect to the device orientation. $\theta_B=0$ ($x$-axis) is defined as the source-drain direction of the Hall bar device. (b) Resistivity as a function of $\theta_B$ for different magnitudes of the in-plane magnetic field, showing a clear two-fold anisotropy. Measurement is taken at $n=\SI{-1.18e12}{\per\centi\meter\squared}$ and $T=\SI{70}{\milli\kelvin}$. (c) Detailed view of the superconducting domes in device B, showing a large and a small superconducting dome on the p-side and n-side of the insulating state. (d) Critical in-plane magnetic field $B_{c\parallel}$ versus temperature along the major and minor axis of the two-fold anisotropy, measured in device A at carrier densities of \SI{-1.44e12}{\per\centi\meter\squared}, \SI{-1.42e12}{\per\centi\meter\squared}, \SI{-1.40e12}{\per\centi\meter\squared} and \SI{-1.23e12}{\per\centi\meter\squared} respectively (from right to left). (e-r) Polar maps of the anisotropic response of the resistivity across the superconducting domes. The carrier densities and temperatures at which (d-q) are measured correlate with the labels in (c). (s) For device B, we extracted the magnitude (represented by height) and the polar angle of the major axis $\theta_\mathrm{maj}$ (represented by the color, see supplementary for definition \cite{supplementary}) of the nematicity at different densities and temperatures, with $|B_\parallel|=\SI{1}{\tesla}$. The data for different temperatures are shifted vertically for clarity. Inside the region shaded in purple, the critical in-plane magnetic field is larger than \SI{1}{\tesla} and cannot be measured in our setup. In the density range of \SIrange{-1.45e12}{-1.2e12}{\per\centi\meter\squared}, the anisotropy polar angle $\theta_\mathrm{maj}$ rotates rapidly with the carrier density, possibly due to the competition with the wedge-like feature we identified in Fig. 2 and Fig. 3. }
\end{figure*}

\subsection{Nematic Superconducting State}

A natural question to ask is whether the superconducting phase exhibits any anisotropic properties as well. To investigate this, we measure the angle-dependent in-plane magnetic field response of the superconducting phase. In magic-angle TBG, the superconductivity is suppressed by an in-plane magnetic field of the same order of magnitude as the Pauli paramagnetic limit \cite{cao2018SC}. Using a vector magnet in a dilution refrigerator, we apply a magnetic field up to \SI{1}{\tesla} in an arbitrary direction within the sample plane (see Fig. 4a for illustration). We compensate for possible sample tilt by applying a small out-of-plane magnetic field, so that the magnetic field is parallel to the sample to within $|B_\perp|<\SI{2}{\milli\tesla}$ at $|B_\parallel|=\SI{1}{\tesla}$ (see Supplementary\cite{supplementary} for detailed calibration procedure). Figure 4b shows an example of the resistivity versus in-plane magnetic field magnitude and angle $\theta_B$ (with respect to the length of the Hall bar, see Fig. 4a). A two-fold anisotropic suppression of the superconductivity can be clearly seen. We have checked that the direction of the current flow is not correlated with the anisotropy direction, and therefore the anisotropic Lorentz force contribution can be excluded.\cite{supplementary} The anisotropy is not aligned with the length or width of the Hall bar either.\cite{supplementary}

The two-fold anisotropy of the in-plane magnetotransport response points towards nematicity that is intrinsic to the superconducting phase, since it breaks the six-fold rotational symmetry of the moir\'e superlattice. We have systematically studied this nematic behavior across the entire superconducting dome of device B. In Figs. 4e-r we show polar maps of the magneto-resistivity at different carrier densities in the hole-doping and electron-doping superconducting domes as labeled in Fig. 4c. At all densities except those in Figs. 4n and r, we find elliptic contours that have major/minor axis ratio up to $~3$. Note that we chose to always measure near $T_c$, since deep inside the superconducting dome the in-plane critical field is usually larger than \SI{1}{\tesla} and cannot be measured in our setup. However, we have confirmed the nematicity in the $T\ll T_c$ region by simultaneously applying a small perpendicular field to partially suppress the superconducting state.\cite{supplementary} At the densities corresponding to Figs. 4n and r, which are outside the superconducting regions, the anisotropy is essentially nonexistent. In device A, we have also observed similar two-fold anisotropic in-plane critical field (Fig. 4d). The critical magnetic field $B_{c\parallel}$ along the major axis extrapolated to zero temperature exceeds that along the minor axis by \SIrange{40}{80}{\percent} in this device.

Fig. 4s shows the evolution of the magnitude and of the director of the nematic component of the superconducting state in device B as a function of carrier density and temperature. Our data shows that the nematic director, as measured by the angle of rotation of the ellipse's major axis, does not appear to be exactly locked to any particular spatial axis, but instead evolves continuously with carrier density. In particular, in the superconducting dome on the hole-doping side of $-n_s/2$, the direction of the major axis varies slowly within \SIrange{-10}{+20}{\degree} in the density range of \SIrange{-1.70e12}{-1.45e12}{\per\centi\meter\squared} (corresponding to the ellipses from Fig. 4e to 4h), while in the range of \SIrange{-1.45e12}{-1.25e12}{\per\centi\meter\squared} (from Fig. 4i-m) the major axis rotates quickly with the carrier density. From Fig. 4i to Fig. 4m, the major axis rotates by $\sim$\SI{90}{\degree}. We note that the latter range of density again coincides with the resistive wedge-like feature for device B, as shown from Figs. 2b and d. The smaller superconducting dome on the electron-doping side near \SI{-1.20e12}{\per\centi\meter\squared} exhibits significant nematicity as well (Fig. 4p and q), with a director pointing from \SI{120}{\degree} to \SI{160}{\degree}. As we explain below, the fact that the nematic director changes direction as a function of doping makes it unlikely that the superconducting anisotropy is simply a response to strain present in the sample. On the contrary, this observation is consistent with spontaneous rotational symmetry-breaking characteristic of an intrinsic nematic superconductor.

\begin{figure*}[!ht]
\includegraphics[width=0.7\textwidth]{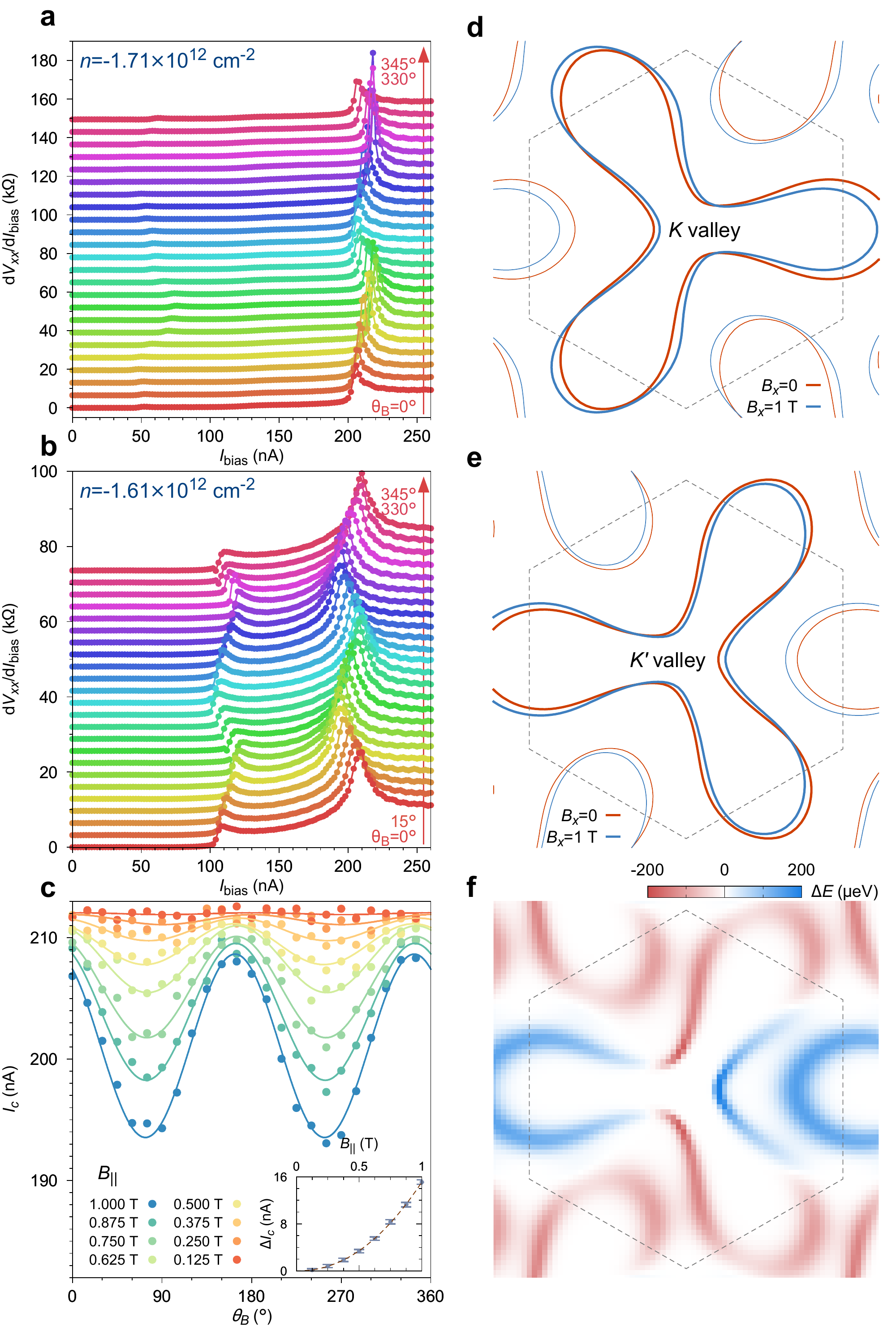}
\caption{\label{fig:fig5}Anisotropic response of the superconducting critical current. (a-b) Differential resistance $dV_\mathrm{xx}/dI_\mathrm{bias}$ versus bias current $I_\mathrm{bias}$ as a function of the orientation of the in-plane magnetic field at two carrier densities. The orientation is indicated by the color, differing by \SI{15}{\degree} between adjacent curves, which are vertically shifted for clarity. (c) Modulation of the larger critical current in (b) by in-plane magnetic fields with different orientations and magnitudes. A sinusoidal function is used to fit the data (see main text). The inset shows the modulation amplitude (peak-peak) as a function of the field magnitude, which can be fit by a power law $\Delta I_c \propto |B_\parallel|^{\alpha}$ with $\alpha\approx 2.1$. (d-e) Calculated Fermi contour of $\theta=\SI{1.09}{\degree}$ TBG at $B_x=0$ and $B_x=\SI{1}{\tesla}$ around $K$ and $K^\prime$ valleys respectively. (f) Energy splitting between states at opposite momentum and opposite valleys along the Fermi surface $\Delta E(\vec{k}) = E_{K'}(\vec{k}) - E_K(-\vec{k})$ at $B_x=\SI{1}{\tesla}$. For comparison, the Zeeman splitting $g\mu_B B$ at $B=\SI{1}{\tesla}$ for $g=2$ is \SI{115}{\micro\electronvolt}. The gray hexagons in (d-f) denote the moir\'{e} Brillouin zone. }
\end{figure*}

\subsection{Anisotropic Response of the Superconducting Gap}

The observation of nematicity puts certain constraints on possible pairing symmetries of the superconducting order parameter.\cite{fernandes2019, kozii2018, isobe2018} One can obtain information about the superconducting gap by measuring the critical current $I_c$. Here, by measuring $I_c$ of device B in the presence of in-plane magnetic fields, we demonstrate that the nematicity is not only manifested in the resistivity measurements, but also creates an anisotropic modulation of the superconducting gap. Figs. 5a-b show the waterfall plots of differential resistance $dV_\mathrm{xx}/dI_\mathrm{bias}$ versus dc bias current $I_\mathrm{bias}$ at two carrier densities, in an in-plane magnetic field $|B_\parallel|=\SI{1}{\tesla}$ along different directions indicated by the colors. At the carrier density in Fig. 5b, the plot shows two critical currents at \SI{110}{\nano\ampere} and \SI{210}{\nano\ampere} respectively, which might be due to domains in the device with different twist angles or nematic directors. Interestingly, at both carrier densities the critical current shows significant two-fold modulation by the in-plane magnetic field direction $\theta_B$. The $\theta_B$ dependence can be fit by a sinusoidal function $\cos2(\theta_B-\theta_{B0})$ (Fig. 5c), where $\theta_{B0}$ is the direction of the major axis.  The modulation amplitude as a function of the in-plane field magnitude is shown in the inset of Fig. 5c, and follows an approximately quadratic power law dependence. 

An anisotropic response in the critical current may originate from (i) the superconducting gap $\Delta$ and/or (ii) anisotropic properties of the underlying normal state ($R_n$). Although we have shown that the normal state exhibits considerable resistance anisotropy at densities near the wedge-like feature in Fig. 3, we argue here that the anisotropic response of the critical current is not a result of the anisotropy of $R_n$. First, Fig. 5a is measured at a density for which there is essentially no resistivity anisotropy in the normal state ($R_{xy}^\mathrm{cr}/R_{xx}=-0.007$ at the lowest $T$ in Fig. 3c), while Fig. 5b is measured at one with significant anisotropy in the normal state ($R_{xy}^\mathrm{cr}/R_{xx}=-0.325$ at the lowest $T$ in Fig. 3c). However, the modulation of the critical current at these two densities shows similar magnitudes. Second, an anisotropy in the resistivity tensor may not necessarily imply a large anisotropic response of the resistivity versus in-plane magnetic field. In fact, as we show in the Supplementary, inside the wedge-like feature in the normal state in device A, we could not measure significant anisotropic response to the in-plane field. Thus, these results suggest that the anisotropic response of the critical current might not be directly related to the resistivity anisotropy of the normal state and hence may originate from an anisotropic superconducting gap.

To discuss the mechanism by which the in-plane field couples to the superconducting gap, we note that if the former couples solely to the spin degree of freedom (and thus the gap is only suppressed by the Zeeman coupling), spin-orbit interaction must be introduced to explain the dependence of $I_c$ on the direction of $B_{\parallel}$. However, the intrinsic spin-orbit coupling in graphene-based systems is known to be very weak. We might consider the following mechanism to reconcile these facts. As illustrated in Fig. 1a, the unit cell of magic-angle TBG has a length scale of $a\sim\SI{14}{\nano\meter}$. Despite the separation between the graphene sheets in TBG being merely $\delta\sim\SI{0.3}{\nano\meter}$, an in-plane magnetic field penetrating them induces a small but non-negligible magnetic flux in the cross-section of the unit cell with an area $S\sim a\cdot \delta$, which modifies the Fermi contours. To demonstrate this effect, we numerically calculated the Fermi contours at $-n_s/2$ for $B_\parallel=0$ and $B_\parallel=\SI{1}{\tesla}$ along the $x$ direction using the Bistritzer-MacDonald continuum model \cite{bistritzer2011}. Figs. 5d and 5e show the original and modified Fermi contours for the $K$ and $K^\prime$ valleys respectively. As can be seen from the contours, a noticeable shift is induced by the in-plane magnetic field. The $K$/$K'$ valley degeneracy is lifted by the momentum shift between the two layers introduced by the in-plane field, which is proportional to $e\delta B_\parallel$, a substantial shift given the small size of the Brillouin zone. If one assumes that only electrons with opposite momentum and valley are allowed to form Cooper pairs in the superconducting phase, the two states from opposite valleys would be at slightly different energies when an in-plane field is applied, which serves to suppress the superconductivity in a similar fashion as the paramagnetic (Zeeman) effect in the case of spins. To more intuitively demonstrate this, Fig. 5f shows the de-pairing energy along the Fermi contour $\Delta E(\vec{k}) = E_{K'}(\vec{k}) - E_K(-\vec{k})$. It is strongly directional dependent and has a similar order of magnitude as the Zeeman energy at $B_\parallel=\SI{1}{\tesla}$ ($g\mu_B B_\parallel\approx\SI{115}{\micro\electronvolt}$ where $g=2$, $\mu_B$ is the Bohr magneton). The de-pairing energy exhibits a  six-fold variation with respect to the direction of the in-plane magnetic field, while the nematic component of the superconducting order can further spontanously break this symmetry down to the observed two-fold symmetry \cite{venderbos2016, kozii2018, fernandes2019_2}. A small strain can further assists to pin down the nematic domain along a given direction.

\begin{figure*}[!ht]
\includegraphics[width=0.6\textwidth]{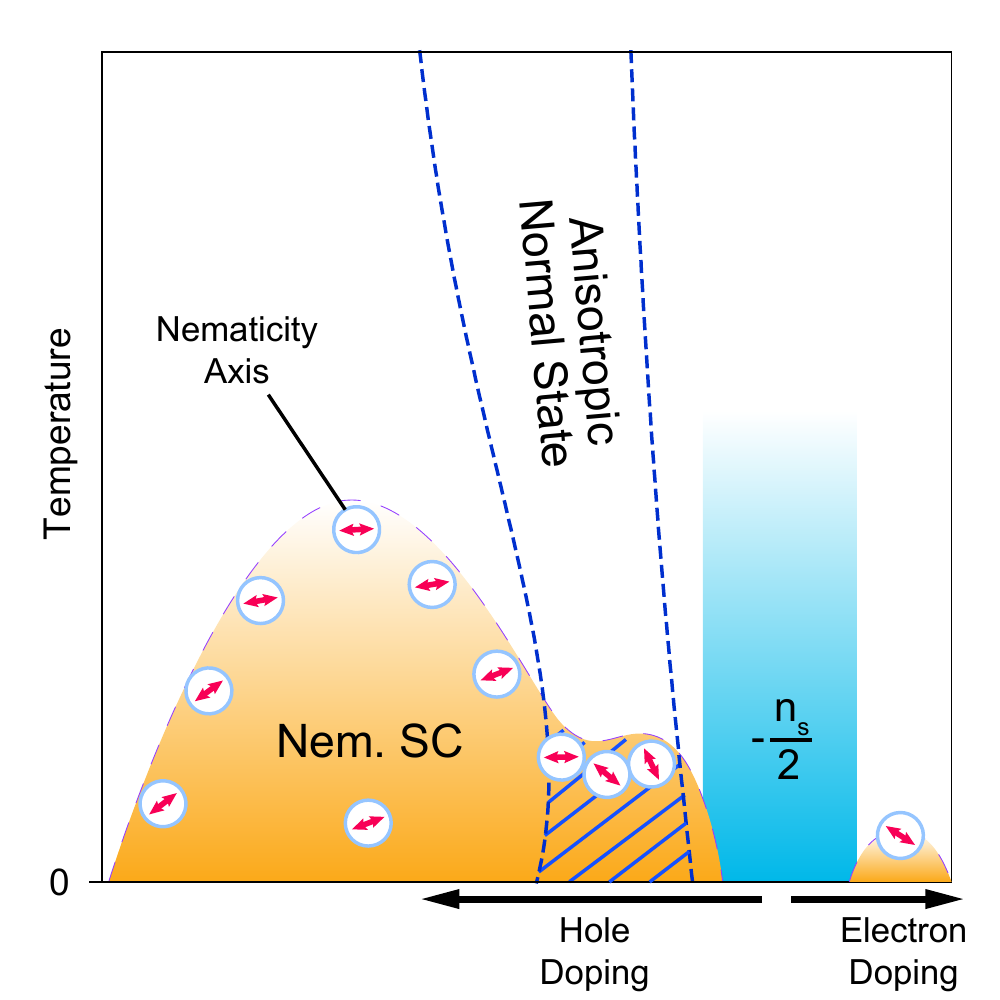}
\caption{\label{fig:fig6} Summary of various competing phases we identified in this article. In the underdoped side of the superconducting dome, we find a normal state anisotropic phase that, at low temperatures, competes with superconductivity, creating a depression in the $T_c$ curve. In the superconducting state we find nematicity, manifested in its response to in-plane magnetic fields. By comparing the extracted nematicity temperature $T_\mathrm{nem}$ to $T_c$ we find that the entire superconducting dome exhibits nematicity, which suggests that the nematicity is intrinsic to the superconductivity and points towards a possible unconventional pairing symmetry. The dashed area denotes the competing region between the two states, which results in a reduction in $T_c$ as well as in a rotation of the nematicity axis. }
\end{figure*}

\section{Discussion}

Our measurements reveal two distinct anisotropic states in the phase diagram of magic-angle TBG: a normal-state wedge-like feature above the superconducting dome and a nematic pairing state. As shown by the longitudinal resistivity and transverse voltage measurements presented in Fig. 2 and Fig. 3, the wedge-like feature is associated with a zero-temperature insulating phase that shows significant resistivity anisotropy, indicative of broken six-fold lattice rotational symmetry. Thus, this normal state phase might be either an electronic nematic state or an electronic smectic state -- i.e. a charge or spin density-wave that, in addition to rotational symmetry, also breaks translational lattice symmetry. In either case, the rotational symmetry-breaking can be described by a two-component 3-state Potts nematic order parameter $\boldsymbol{\Phi} = \Phi \left( \cos 2\theta_n, \sin 2\theta_n \right)$, with $\theta_n$ restricted to three possible values \cite{hecker2018,fernandes2019_2}. Electronic correlations might be important for the formation of such a state. Twisted bilayer graphene is well-known to exhibit van Hove singularities (vHs), which in general do not occur exactly at half-filling \cite{li2010, luican2011}. Near the vHs, it has been theoretically shown that the significant nesting between the $K$- and $K'$-valley Fermi contours might induce density wave ordering \cite{isobe2018}. Remarkably, recent scanning tunneling experiments have identified prominent rotational-symmetry-broken features in the normal state local density of states,\cite{kerelsky2019, jiang2019, li2017_2}. Alternatively, strong-coupling models can also yield nematic and density-wave states \cite{vafek2019,dodaro2018}. Importantly, the fact that only one dominant nematic domain is observed implies the existence of some small residual strain in the device, which selects that particular domain. One can rule out the scenario in which the anisotropic state itself is a trivial consequence of such a strain, because the wedge-like feature is restricted to narrow temperature and doping ranges. In contrast, strain-induced anisotropy should persist at all temperatures and over a much wider doping range.

For the superconducting phase, its remains to be seen whether its nematic character, as revealed by the measured in-plane anisotropy of the critical field, can be reconciled with $s$-wave pairing. On the other hand, it may be more naturally explained in terms of a two-component $p$-wave/$d$-wave gap of the form $\boldsymbol{\Delta} = \Delta \left( \cos \theta_s, \sin \theta_s \right)$, indicative of an unconventional pairing mechanism \cite{chubukov2019}. Here, the parameter $\theta_s$, responsible for the breaking of the six-fold rotational symmetry, is associated with the orientation of the $H_c$ ellipses in Fig. 4 (Ref.\cite{venderbos2016}). The fact that only one orientation is observed for a given doping suggests that strain is pinning it. However, because the ellipse orientation rotates continuously over the doping range \SIrange{-1.45e12}{-1.25e12}{\per\centi\meter\squared}, we can conclude that the anisotropy of the superconducting state is intrinsic, i.e. it would be present even for zero strain. To see this, we follow Ref. \cite{fernandes2019_2} and note that, to lowest order in a free-energy expansion, uniaxial strain $\varepsilon$ couples to the superconducting order parameter as $\varepsilon \Delta^2 \cos(2\theta_s - 2 \alpha)$, where $\alpha$ is the direction strain is applied. This term alone only allows two possible relative orientations between $\theta_s$ and $\alpha$, namely, \SI{0}{\degree} and \SI{90}{\degree}, depending on the sign of $\varepsilon$ (i.e. compressive or tensile strain). Thus, because for a given device $\alpha$ is presumably fixed, we would expect the same ellipse orientation for all doping levels. However, if $\theta_s$ breaks the rotational symmetry on its own (i.e without strain), the free energy has another relevant term $\Delta^6 \cos 6\theta_s$. In the absence of strain, this term fixes $\theta_s$ to three values (modulo $\pi$). When combined with the strain-coupling term of the free energy, it allows $\theta_s$ to continuously rotate within a range of values, which depend on phenomenological parameters. While a more detailed analysis is presented in the supplementary material, the simple fact that the ellipses orientations are not the same for all doping levels provides strong evidence that the nematic superconductivity is an intrinsic property of magic-angle TBG. Indeed, nematicity is observed essentially across the entire superconducting dome. The intrinsic inhomogeneity of TBG devices, as manifested for instance in twist angle variations across the sample, suggests that the strain that pins the nematic director may also be inhomogeneous. While further studies are needed, we note that such an inhomogeneous strain would act as a random field to the Potts-nematic order parameter, which can strongly affect the nematic properties in 2D \cite{blankschtain1984}.

The various phases discussed throughout this article are summarized in Fig. 6. The fact that an anisotropic response to an in-plane magnetic field is seen only in the superconducting state, but not in the wedge-like feature, suggests that the origins of nematicity in the normal and superconducting states are likely different. This is also consistent with the fact that these two orders compete, as evident from the suppression of $T_c$ when the wedge-like feature intersects with the superconducting dome. However, since both phases break the same six-fold lattice rotational symmetry, the order parameters of these two phases can interact beyond mere competition, which may be responsible for the rapid change of the ellipse direction in the coexisting region of the phase diagram (see Supplementary Material \cite{supplementary}). Moreover, normal-state nematic fluctuations may play an important role in favoring a superconducting ground state that is also nematic. While the onset of nematicity and of superconductivity seem very close in our experiment (see Supplementary Material), it is possible that the nematic order in magic-angle TBG persists even above $T_c$, a phenomenon known as vestigial nematic order \cite{nie2014,fernandes2019,hecker2018}. Interestingly, in Fig. 3b, there is a region just above the superconducting dome with positive transverse voltage signal at $n\approx\SI{-1.65e12}{\per\centi\meter\squared}$ and $T\approx \SI{2}{\kelvin}$. This not only has opposite sign than the anisotropy of the wedge-like state, but it also disappears when superconductivity is suppressed (Fig. 3c). Thus, this feature might be explained by a vestigial nematic order that forms prior to the condensation of Cooper pairs.\cite{nie2014,fernandes2019, hecker2018}. Scanning probe experiments are encouraged in the future to confirm this nematic phase above the superconducting transition.

In summary, our experiments extend the already rich phase diagram of magic-angle TBG to include a nematic superconducting state and an anisotropic normal state above the `underdoped' part of the superconducting dome. The competition between them results in a reduction of $T_c$ and in a fast rotation of the nematic director of the superconducting state. Our results pioneer the study of competing/intertwined quantum phases in a highly tunable two-dimensional correlated platform, which in turn may shed more light onto the unconventional superconductivity in iron-based compounds, doped Bi\textsubscript{2}Se\textsubscript{3} and other nematic superconductors.

\begin{acknowledgments}
We acknowledge helpful discussions with P. A. Lee, S. Todadri, A. Vishwanath, A. Hristov, I. Fisher, J. Venderbos, and S. A. Kivelson. 
\end{acknowledgments}

\end{document}